\documentclass[a4paper,conference,onecolumn]{IEEEtran}

\usepackage[utf8]{inputenc}
\usepackage[T1]{fontenc}
\usepackage{url}
\usepackage{ifthen}
\usepackage{amsfonts}
\usepackage{cite}
\usepackage{tikz}
\usepackage{cancel}
\usepackage{xcolor}
\usepackage{float}
\usepackage{bm}
\usepackage{scalerel}

\usepackage{amsthm}

\newtheorem{theorem}{Theorem}

\newtheorem{lemma}{Lemma}

\theoremstyle{definition}
\newtheorem{definition}{Definition}
\usepackage{amssymb}

\usetikzlibrary{patterns}
\usetikzlibrary{shapes,arrows}
\newcommand{\Exp}[2][]{\ensuremath{\mathbb{E}_{#1}\left[#2 \right]}} 

\newcommand{\eqann}[2][=]{\overset{\mathclap{(\text{#2})}}{#1}} 
\newcommand{\eqannref}[1]{$(\text{#1})$}
\newcommand{\bv}[1]{\mathbf{#1}} 
\newcommand{\rv}[1]{\mathsf{#1}} 
\newcommand{\Prob}[1]{\ensuremath{\mathbb{P} \left(#1 \right)}} 

\newcounter{tempEquationCounter} 
\newcounter{thisEquationNumber}

\usepackage[cmex10]{mathtools} 
\begin{document}
\title{Broadcasting Information subject to State Masking over a MIMO State Dependent Gaussian Channel }



 \author{%
   \IEEEauthorblockN{Michael Dikshtein\IEEEauthorrefmark{1},
     Anelia Somekh-Baruch \IEEEauthorrefmark{2}
     and Shlomo Shamai (Shitz)\IEEEauthorrefmark{1}}
   \IEEEauthorblockA{\IEEEauthorrefmark{1}%
              Department of EE, 
              Technion, 
              Haifa 32000, Israel,
              \{michaeldic@campus,sshlomo@ee\}.technion.ac.il}
   \IEEEauthorblockA{\IEEEauthorrefmark{2}%
              Faculty of Engineering, Bar-Ilan University, 
              Ramat-Gan, 52900, Israel,
              anelia.somekhbaruch@gmail.com}
 }
%

\maketitle

\begin{abstract}
	The problem of channel coding over the Gaussian multiple-input multiple-output (MIMO) broadcast channel (BC) with additive independent Gaussian states is considered. The states are known in a noncausal manner to the encoder, and it wishes to minimize the amount of information that the receivers can learn from the channel outputs about the state sequence. The state leakage rate is measured as a normalized blockwise mutual information between the state sequence and the channel outputs' sequences. We employ a new version of a state-dependent extremal inequality and show that Gaussian input maximizes the state-dependent version of Marton's outer bound. Further we show that our inner bound coincides with the outer bound. Our result generalizes previously studied scalar Gaussian BC with state and MIMO BC without state.
\end{abstract}


\section{Introduction}

\begin{figure}
	\centering
	\tikzstyle{block} = [draw, rectangle, minimum width = 0.5cm, minimum height = 0.5cm]
	\tikzstyle{sum} = [draw,circle]
	\tikzstyle{input} = [coordinate]
	\tikzstyle{dummy} = [coordinate]
	\tikzstyle{output} = [coordinate]
	\tikzstyle{amp} = [draw,shape border rotate = 180,regular polygon,regular polygon sides=3]
	
	\begin{tikzpicture}
	
	\node [input]	(in)	at 	(0,0) 		{};
	\node [block]	(enc) 	at 	(1,0)		{\tiny{Enc}};
	\node [block]	(state) at 	(2.5,1.5)		{\tiny{$ P_{\rv{S}} $}};
	\node [draw, rectangle, minimum width = 1cm, minimum height = 1.5cm]	(ch)	at 	(2.5,0)		{\tiny{$P_{\rv{Y}_1 \rv{Y}_2|\rv{X} \rv{S}}$}};
	\node [block]	(dec1)	at	(4,0.5)		{\tiny{Dec 1}};
	\node [block]	(dec2)	at	(4,-0.5)		{\tiny{Dec 2}};
	\node [output]	(out1)	at	(5,0.5)		{};
	\node [output]	(out2)	at	(5,-0.5)		{};
	
	\draw [->] (in) node[left] {\tiny{$\rv{M}_0, \rv{M}_1, \rv{M}_2$}} --  (enc);
	\draw [->] (state) -- (state-|enc) -- node [right] {\tiny{$ \rv{S}^n $}} (enc) ;
	\draw [->] (state) -- node [right] {\tiny{$ \rv{S}^n $}} (ch);
	\draw [->] (enc) -- node[above] {\tiny{$\rv{X}^n$}} (ch);
	\draw [->] (ch.40) --  node[above] {\tiny{$\rv{Y}_1^n$}} (dec1)  ;
	\draw [->] (ch.320) --  node[above] {\tiny{$\rv{Y}_2^n$}} (dec2)  ;
	\draw [->] (dec1) --  (out1);
	\draw [->] (dec2) --  (out2);
	\node at (5.5,0.7) {\tiny{$\hat{\rv{M}}_0,\hat{\rv{M}}_1$}};
	\node at (6,0.3) {\tiny{$E_{1}\geq \frac{1}{n} I(\rv{S}^n;\rv{Y}_1^n)$}};
	\node at (5.5,-0.3) {\tiny{$\hat{\rv{M}}_0,\hat{\rv{M}}_2$}};
	\node at (6,-0.7) {\tiny{$E_{2}\geq \frac{1}{n} I(\rv{S}^n;\rv{Y}_2^n)$}};
	\end{tikzpicture}
	\caption{System model for general BC subject to state masking constraints.}
	\label{figure:system_model}
\end{figure}
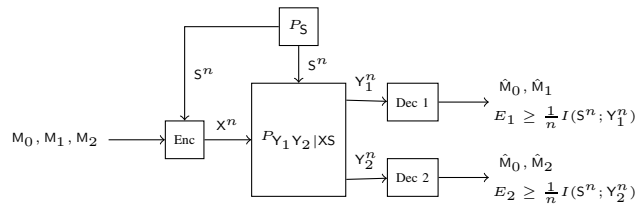

We consider the problem of reliable transmission of pure digital information over a two-user MIMO Gaussian BC with an additive interference, modeled as state, which is known in a noncausal manner to the encoder. In our setting, we impose an additional requirement to reduce the knowledge of the receivers regarding the state, measured as a normalized blockwise mutual information between the state sequence and the received sequences, as depicted in Figure \ref{figure:system_model}. The problem under consideration can act as a simplified model to many evolving practical communication systems. Consider, for example, a base station (BS) which is equipped with multiple antennas while the cell is partitioned to various sectors. The BS wishes to prevent leakage of information between the sectors, that is, to minimize the knowledge of mobile users in a specific sector regarding the messages intended for other sectors. In such a scenario, the part of the BS signal intended for other sectors is modeled as a state sequence. The state sequence is known to the BS in a noncausal manner since it is the one that generates it.

Problems of information transmission over channels with a noncausal channel state information (CSI) have been the subject for extensive study. The single-letter expression for the capacity of the point-to-point discrete memoryless channel (DMC) with noncausal CSI at the encoder (the G-P channel) was derived in the seminal work of Gel'fand and Pinsker \cite{gelfand1980coding}. One of the most interesting special cases of the G-P channel is the Gaussian additive noise and interference setting in which the additive interference plays the role of the state sequence, which is known non-causally to the transmitter. Costa showed in \cite{costa1983wdp} that the capacity of this channel is equal to the capacity of the same channel without additive interference. The capacity achieving scheme of \cite{costa1983wdp} (which is that of \cite{gelfand1980coding} applied to the Gaussian case), is termed ``writing on dirty paper" (WDP). Cohen and Lapidoth \cite{cohen2002gwdp} showed that any interference sequence can be totally removed when the channel noise is ergodic and Gaussian.

The general DM-BC was introduced by Cover \cite{cover1972broadcast}. The capacity region of the DM-BC is still an open problem. The largest known inner bound on the capacity region of the DM-BC was derived by Marton \cite{marton1979coding}. The best outer bound for DM-BC with common and private messages is due to Nair and El Gamal \cite{nair2007outer}. There are, however, some special cases in which the capacity region is fully characterized. For example, the capacity region of the degraded DM-BC was established by Gallager \cite{gallager1974capacity}. The capacity region of the Gaussian BC was derived by Bergmans \cite{bergmans1974}. Bergmans established the converse result using the conditional version of the Entropy Power Inequality (EPI). It can be shown \cite{gardner2002brunn} that the Strengthened Young's Inequality implies the EPI.  An interesting result is the capacity region of the Gaussian MIMO BC which was established by Weingarten et al. \cite{weingarten2006capacity}. The authors showed that Bergmans technique cannot be directly applied to the MIMO scenario since a certain proportionality condition is not always satisfied, hence they introduced a new notion of \textit{an enhanced channel} and used it jointly with the EPI to show their result. The capacity achieving scheme relies on the dirty paper coding technique.  Liu and Viswanath \cite{liu2007extremal} developed \textit{an extremal inequality} proof technique and showed that it can be used to establish a converse result in various Vector Gaussian  multiterminal networks, including the Gaussian MIMO BC with private messages. Recently, Geng and Nair \cite{geng2014capacity} employed the new factorization of concave envelope technique to characterize the capacity region of Gaussian MIMO BC with common and private messages.

Degraded DM-BC with causal and noncausal CSI was introduced by Steinberg \cite{steinberg2005coding}. Inner and outer bounds were derived on the capacity region. For the special case in which the nondegraded user is informed about the channel parameters, it was shown that the bounds are tight, thus deriving the capacity region for that case. The general DM-BC with a noncausal CSI at the encoder was studied by Steinberg and Shamai \cite{steinberg2005achievable}. An inner bound was derived, and it was shown to be tight for the Gaussian BC with independent additive interference at both channels. Outer bounds for DM-BC with CSI at the encoder were derived in \cite{khosravi2011capacity}.

The problem of state-masking and information rate trade-off was introduced in \cite{merhav2007information}. In that work, the state sequence was treated as an undesired information that leaks to the receiver and is known to the transmitter. The measure of ability of the receiver to learn about the state from the received sequence was defined as the normalized block-wise mutual information between the state sequence $ \rv{S}^n $ and the received sequence $ \rv{Y}^n $, that is, $ I(\rv{S}^n;\rv{Y}^n)/n$.

The concept of state amplification is a dual problem to state masking. Kim et al. \cite{kim2008state} considered the problem of transmitting data at rate $ R $ over a DMC with random parameters and CSI at the encoder and simultaneously conveying the information about the channel state itself to the receiver.  They defined the channel state uncertainty reduction rate to be $ \Delta \triangleq \frac{1}{n} (H(S^n)-\log |L_n|) $, where $ |L_n| $ is the receiver list size in list decoding of the state, and found the $ (R,\Delta) $ achievable region.

Courtade \cite{courtade2012information} considered a joint scenario, with two-encoder source coding setting where one source is to be amplified, while the other source is to be masked. Koyluoglu et al. \cite{koyluoglu2016state} considered a state-dependent BC with state sequence known in a noncausal manner to Alice (the transmitter), and its goal is to effectively convey the state to Bob (receiver 1) while "masking" it from Eve (receiver 2). Liu and Chen \cite{liu2009message} considered the problem of message transmission and state estimation over the Gaussian BC, where both received signals interfered by same additive Gaussian state.
Grover and Sahai \cite{grover2010witsenhausen} related the problem of state masking to Witsenhausen's Counter-example \cite{witsenhausen1968counterexample}. Tutuncuoglu et al. \cite{tutuncuoglu2014state} studied the problem of state amplification and state masking in an energy harvesting binary channel, where the energy source is modeled as state. A privacy-constrained information extraction problem was recently considered by Asoodeh et al. \cite{asoodeh2016}. A good tutorial on channel coding in the presence of CSI that also covers the state masking setting can be found in \cite{keshet2008channel}.

In our previous work \cite{dikshtein2018masking}, we extended the state masking scenario to the state-dependent DM-BC with noncausal CSI at the encoder.  We developed inner and outer bounds and showed that they are tight for a special case of zero-rates transmission and the scalar Gaussian BC with additive state. Our main goal in this work is to address the MIMO Gaussian BC with additive Gaussian state known to the transmitter in a noncausal manner, for which we show that a new optimization problem should be solved. We develop a new, conditional, extremal inequality,  and show that a Gaussian input distribution solves the optimization problem mentioned above. Our approach to evaluate the inner-bound, also introduces some novelty on how to properly choose the optimal coefficients, which can contribute to solve other multi-terminal Gaussian problems.

\section{Notations and Problem Formulation}

Throughout the paper, random variables are denoted using a sans-serif font, e.g., $ \rv{X} $, their realizations are denoted by the respective lower case letters, e.g., $ x $, and their alphabets are denoted by the respective calligraphic letter, e.g., $ \mathcal{X} $. Let $ \mathcal{X}^n $ stand for the set of all $ n $-tuples of elements from $ \mathcal{X} $. An element from $ \mathcal{X}^n $ is denoted by $ x^n = (x_1,x_2,\dots , x_n) $. The probability distribution function of $ \rv{X} $, the joint distribution function of $ \rv{X} $ and $ \rv{Y} $, and the conditional distribution of $ \rv{X} $ given $ \rv{Y} $ are denoted by $ P_\rv{X} $, $ P_{\rv{X},\rv{Y}} $ and $ P_{\rv{X}|\rv{Y}} $ respectively. The expectation of $ \rv{X} $ is denoted by $ \Exp{\rv{X}} $. The cross-covariance matrix of two random vectors $ \bv{X} $ and $ \bv{Y} $ is denoted as $ \Sigma_{ \bv{X}\bv{Y}} \triangleq \Exp{\bv{X}\bv{Y}^T} $. The probability of an event $ \mathcal{E} $ is denoted as $ \Prob{\mathcal{E}} $. A set of consecutive integers starting at $ 1 $ and ending in $ 2^{nR} $ is denoted as $ \mathcal{I}^{(\! n \!)}_{R} \triangleq \{1,2,\dots, 2^{nR} \} $.

An $(2^{nR_1},2^{nR_2},n) $ code for the broadcast channel with state sequence $ \rv{S}^n $ known non-causally at the encoder consists of
\begin{itemize}
	\item two message sets $ \mathcal{I}^{(\! n \!)}_{R_1}  $ and $ \mathcal{I}^{(\! n \!)}_{R_2} $,
	\item an encoder that assigns a codeword $ x^n(m_1,m_2,s^n) $ to each message-state triple $ (m_1,m_2,s^n) \in  \mathcal{I}^{(\! n \!)}_{R_1} \times \mathcal{I}^{(\! n \!)}_{R_2} \times \mathcal{S}^n $, 
	\item two decoders, where decoder 1 assigns an estimate   and $ \hat{m}_1 \in \mathcal{I}^{(\! n \!)}_{R_1} $ to each received sequence $ y_1^n $, and decoder 2 assigns an estimate  $ \hat{m_2} \in \mathcal{I}^{(\! n \!)}_{R_2} $ to each received sequence $ y_2^n $.
\end{itemize}
Let $\hat{\rv{M}}_1 $ and $\hat{\rv{M}}_2$ denote the outputs of decoder $ 1 $ and decoder $ 2 $, respectively.  We assume that the message pair $ ( \rv{M}_1, \rv{M}_2) $ is uniformly distributed over $  \mathcal{I}^{(\! n \!)}_{R_1} \times \mathcal{I}^{(\! n \!)}_{R_2} $. The average probability of error is defined as
\begin{equation}
	P_e^{(n)} =	\Prob{\bigcup_{k=1}^2 \{\hat{\rv{M}}_{k} \neq \rv{M}_k \} }.
\end{equation}

We are interested in the interplay between reliable coding at rate pairs $ (R_1,R_2) $ which we would like to keep as high as possible and the (normalized) mutual informations $ I(\bv{S}^n;\bv{Y}_1^n)/n $ and $ I(\bv{S}^n;\bv{Y}_2^n)/n $, which we would like to make as small as possible. 
\begin{definition}
	
For a given covariance matrix $ K $, a quadruple $ (R_1,R_2,E_{1},E_{2}) $ is said to be achievable if for every $ \epsilon>0 $ and sufficiently large $ n $, there exists a sequence of $ (2^{nR_1},2^{nR_2},n) $ codes such that the following conditions are simultaneously satisfied:
\begin{subequations}
	\begin{align}
	\frac{1}{n} \sum_{i=1}^{n} \bv{X}_i\bv{X}_i^T &\preceq K ,\\
	P_e^{(n)} &\leq \epsilon,\\
 	\frac{1}{n} I(\bv{S}^n;\bv{Y}_k^n)  &\leq E_{k}+\epsilon, \quad  k=1,2.
 	\end{align}
\end{subequations}
\end{definition}
\begin{definition}
	The achievable region $ \mathcal{R} $ is the closure of the set of all achievable quadruples $ (R_1,R_2,E_{1},E_{2}) $.

\end{definition}

\section{Preliminaries}
We use inner and outer bounds that were derived in \cite{dikshtein2018masking} for the general DM-BC with random parameters and particularly utilize the private messages only case by setting $ \rv{W} = \emptyset $.
\begin{lemma}[{Proposition 1 in {\cite{dikshtein2018masking}}}] \label{lemma:inner_bound_dmbc}
	An achievable region $ \mathcal{R} $ consists of a quadruple $ (R_1,R_2,E_1,E_2) $ that satisfies the following conditions
	\begin{subequations} \label{eq:SDDMBC_inner_bound}
		\begin{align}
		R_1 &\leq I(\rv{U};\bv{Y}_1)-I(\rv{U};\bv{S}),\\
		R_2 &\leq I(\rv{V};\bv{Y}_2)-I(\rv{V};\bv{S}),\\
		R_1+R_2 &\leq I(\rv{U};\bv{Y}_1)-I(\rv{U};\bv{S})  + I(\rv{V};\bv{Y}_2)-I(\rv{V};\bv{S}) - I(\rv{U};\rv{V}|\bv{S}) ,\\
		E_1 &\leq I(\bv{S};\rv{U},\bv{Y}_1),\\
		E_2 &\leq  I(\bv{S};\rv{V},\bv{Y}_2),
		\end{align}
	\end{subequations}
	for some pdf $ P_{\bv{S}\rv{U}\bv{X}\bv{Y}_1\bv{Y}_2}=P_{\bv{S}}P_{\rv{U}\rv{V}\bv{X}|\bv{S}} P_{\bv{Y}_1\bv{Y}_2|\bv{X}\bv{S}} $.
\end{lemma}
The main idea behind the proof of the inner bound is an integration of the Marton and the G-P coding schemes, where for each message, a subcodebook is generated, whose size is large enough such that for every state sequence $ s^n $, a jointly typical auxiliary codeword can be found in the subcodebook.

Next, we provide the outer bound on $ \mathcal{R} $.
\begin{lemma}[Proposition 2 in {\cite{dikshtein2018masking}}] \label{lemma:outer_bound_dmbc}
	If a rate quadruple $ (R_1,R_2,E_1,E_2) $ is achievable for the DM-BC with random parameters and CSI known noncausally at the transmitter, then there exists a distribution $P_{\rv{U}\rv{V}\bv{X}|\bv{S}}$ such that the following inequalities are satisfied:
	\begin{subequations} \label{eq:SDDMBC_outer_bound}
		\begin{align} \label{eq:MIMO_Gaussian_Private_R1}
		R_1 &\leq I(\rv{U};\bv{Y}_1|\bv{S}),\\
		\label{eq:MIMO_Gaussian_Private_R2}
		R_2 &\leq I(\rv{V};\bv{Y}_2|\bv{S}),\\
		R_1+R_2 &\leq  I(\rv{U};\bv{Y}_1|\bv{S})+I(\bv{X};\bv{Y}_2|\rv{U},\bv{S}),\\ \label{eq:SDDMBS_sum_rate_upper_bound2}
		R_1+R_2 &\leq I(\bv{X};\bv{Y}_1|\rv{V},\bv{S})+I(\rv{V};\bv{Y}_2|\bv{S}),\\
		E_k &\geq I(\bv{S};\bv{Y}_k) \quad k=1,2,
		\end{align}
	\end{subequations}
	where $P_{\bv{S}\rv{U}\rv{V}\bv{X}\bv{Y}_1\bv{Y}_2}=P_{\bv{S}} P_{\rv{U}\rv{V}\bv{X}|\bv{S}}P_{\bv{Y}_1\bv{Y}_2|\bv{X}\bv{S}}$.
\end{lemma}


\section{MIMO Gaussian Broadcast Channel}
Our goal in this paper is to characterize the achievable region for the MIMO Gaussian State-Dependent BC with masking constraints. We show that a Gaussian input distribution maximizes the outer bound in Lemma \ref{lemma:outer_bound_dmbc}. In order to show this a new proof technique is needed and we show the motivation to develop such technique. 

The general two-user MIMO Gaussian BC with state \cite{weingarten2006capacity}, is an additive interference and noise channel where each time sample can be represented using the following equations:
\begin{align}
\bv{Y}_1 = \bv{X}+\bv{S}_1+\bv{Z}_1, \\
\bv{Y}_2 = \bv{X}+\bv{S}_2+\bv{Z}_2,
\end{align}
where $ \bv{X} $, $ \bv{S}_1 $, $ \bv{S}_2 $, $ \bv{Z}_1 $, $ \bv{Z}_2 $ are all real vectors of size $ t \times 1 $ and
\begin{itemize}
	\item $ \bv{X} $ is the input vector whose covariance matrix satisfies $ \Exp{\bv{X} \bv{X}^T}  \preceq K$ for some $ K \succeq 0 $,
	\item $ \bv{Y}_k $ is the output vector, $ k \in \{1,2\} $,
	\item $ \bv{S}_k $ is a real Gaussian random vector with zero mean and a covariance matrix $ K_{\bv{S}_k} = \Exp{\bv{S}_k \bv{S}_k^T} \succeq 0 $,
	\item $ \bv{Z}_k $ is a real Gaussian random vector with zero mean and a covariance matrix $ K_{\bv{Z}_k} = \Exp{\bv{Z}_k \bv{Z}_k^T} \succeq 0 $.
\end{itemize}

In the following we evaluate the bounds from Lemma \ref{lemma:inner_bound_dmbc} and Lemma \ref{lemma:outer_bound_dmbc} for the MIMO Gaussian BC setting. We first state this result in the following theorem. 

Let,
\begin{equation} \label{eq:K_S_definition}
K_{\bv{S}} = \begin{pmatrix}
K_{\bv{S}_1} & 0 \\
0 & K_{\bv{S}_2}
\end{pmatrix}.
\end{equation}
\begin{theorem} \label{theorem:achivable_region_mimo_bc_private}
	A rate-leakage region of the MIMO Gaussian State-Dependent BC with private messages is the quadruple $ (R_1,R_2,E_{1},E_{2}) $ such that
	\begin{align}
	R_1 &\leq  \frac{1}{2}\log \frac{| K_{\bv{X}_1} +K_{\bv{Z}_1}|}{|K_{\bv{Z}_1}|}, \label{eq:mimo_bc_r1}\\
	R_2 &\leq \frac{1}{2}\log \frac{|K-\Sigma_{\bv{X}\bv{S}} K_\bv{S}^{-1} \Sigma_{\bv{X}\bv{S}}^T+K_{\bv{Z}_2}|}{|K_{\bv{X}_1} +K_{\bv{Z}_2}|} \label{eq:mimo_bc_r2},\\
	E_{1} &= \frac{1}{2} \log \frac{|K+\Sigma_{\bv{X} \bv{S}_1}+\Sigma_{\bv{X} \bv{S}_1 }^T+K_{\bv{S}_1}+K_{\bv{Z}_1}|} {|K- \Sigma_{\bv{X} \bv{S}}  K_{\bv{S}}^{-1} \Sigma_{\bv{X}\bv{S}}^T +K_{\bv{Z}_1} |},\label{eq:mimo_bc_e1}\\
	E_{2} &= \frac{1}{2} \log \frac{|K+\Sigma_{\bv{X} \bv{S}_2}+\Sigma_{\bv{X} \bv{S}_2 }^T+K_{\bv{S}_2}+K_{\bv{Z}_2}|} {|K- \Sigma_{\bv{X} \bv{S}}  K_{\bv{S}}^{-1} \Sigma_{\bv{X}\bv{S}}^T +K_{\bv{Z}_2} |},\label{eq:mimo_bc_e2}
	\end{align}
	for some covariance matrices $ (K_{\bv{X}_1},\Sigma_{ \bv{X}\bv{S}_1},\Sigma_{ \bv{X}\bv{S}_2}) $, such that  $ 0 \preceq K_{\bv{X}_1} \preceq K-\Sigma_{\bv{X}\bv{S}}K_{\bv{S}}^{-1}\Sigma_{\bv{X}\bv{S}}^T $, where
	\begin{equation} \label{eq:Sigma_X_S_definition}
	\Sigma_{\bv{X}\bv{S}} = \begin{pmatrix}
	\Sigma_{\bv{X}\bv{S}_1} & \Sigma_{\bv{X}\bv{S}_2}
	\end{pmatrix}.
	\end{equation}
\end{theorem}
The information rate region in \eqref{eq:mimo_bc_r1} and \eqref{eq:mimo_bc_r2} is similar to the MIMO BC without state in \cite{weingarten2006capacity}. The main difference is that part of the transmitted signal, reflected by the covariance matrix $ \Sigma_{\bv{X}\bv{S}} $, is utilized to mask the state sequence.

\section{Proof of Theorem \ref{theorem:achivable_region_mimo_bc_private}}
Denote $ \bv{S} =(\bv{S}_{1}^T,\bv{S}_{2}^T)^T $.
\subsection{Proof of the converse part of Theorem \ref{theorem:achivable_region_mimo_bc_private}}
Consider the RHS of \eqref{eq:MIMO_Gaussian_Private_R2}, applied to the vector case
\begin{align*}
I(\rv{V};\bv{Y}_2|\bv{S}) &= I(\rv{V},\bv{X};\bv{Y}_2|\bv{S})-I(\bv{X};\bv{Y}_2|\rv{V},\bv{S})\\
&\eqann[=]{a} I(\bv{X};\bv{Y}_2|\bv{S})-I(\bv{X};\bv{Y}_2|\rv{V},\bv{S}),
\end{align*}
where \eqannref{a} follows since $ \bv{Y}_2=\bv{X}+\bv{S}_2+\bv{Z}_2 $ and hence $ h(\bv{Y}_2|\rv{V},\bv{X},\bv{S})= h(\bv{Z}_2) = h(\bv{Y}_2|\bv{X},\bv{S})$. Thus the weighted sum rate upper bound of $ R_1+\mu R_2 $ can be written as
\begin{align*}
&R_1+\mu R_2 \\
&\leq \mu  I(\bv{X};\bv{Y}_2|\bv{S})+ I(\bv{X};\bv{Y}_1|\rv{V},\bv{S})-\mu I(\bv{X};\bv{Y}_2|\rv{V},\bv{S}) \\
&\leq \sup_{
	\substack{
	P_{\scaleto{\rv{V}\bv{X}|\bv{S}}{4pt}}:
	\rv{V}\rightarrow (\bv{X},\bv{S})\rightarrow (\bv{Y}_1,\bv{Y}_2) \\
	\Exp{\bv{X} \bv{X}^T} \preceq K
	}
} 
\mu  I(\bv{X};\bv{Y}_2|\bv{S})+ I(\bv{X};\bv{Y}_1|\rv{V},\bv{S})
-\mu I(\bv{X};\bv{Y}_2|\rv{V},\bv{S})  \\
&\leq \sup_{
	P_{\scaleto{\bv{X}|\bv{S}}{4pt}}: \Exp{\bv{X} \bv{X}^T} \preceq K
} 
\mu  I(\bv{X};\bv{Y}_2|\bv{S}) + \sup_{
	\substack{
		P_{\scaleto{\rv{V}\bv{X}|\bv{S}}{4pt}}:
		\rv{V}\rightarrow (\bv{X},\bv{S})\rightarrow (\bv{Y}_1,\bv{Y}_2) \\
		\Exp{\bv{X} \bv{X}^T} \preceq K
	}
} 
I(\bv{X};\bv{Y}_1|\rv{V},\bv{S})-\mu I(\bv{X};\bv{Y}_2|\rv{V},\bv{S}) ,
\end{align*}
where $ \mu > 1 $.

Consider the term
\begin{equation} \label{eq:mutual_iformation_X_Y2_given_S}
I(\bv{X};\bv{Y}_2|\bv{S}) = h(\bv{X}+\bv{Z}_2|\bv{S})-h(\bv{Z}_2).
\end{equation}
The first term in the RHS of \eqref{eq:mutual_iformation_X_Y2_given_S} can be upper bounded as
\begin{equation} \label{eq:maximum_entropy_XpZ2_given_S}
h(\bv{X}+\bv{Z}_2|\bv{S}) 	\leq \frac{1}{2} \log (2\pi e)^t \left| K  - \Sigma_{\bv{X}\bv{S}} K_{\bv{S}}^{-1}\Sigma_{\bv{X}\bv{S}}^T+K_{\bv{Z}_2} \right|,
\end{equation}
and the second term is simply
\begin{equation} \label{eq:entropy_Z2}
h(\bv{Z_2}) = \frac{1}{2} \log (2\pi e)^t \lvert K_{\bv{Z}_2} \rvert.
\end{equation}
As for the difference between mutual informations, we obtain
\begin{equation} \label{eq:difference_of_mutual_informations}
I(\bv{X};\bv{Y}_1|\rv{V},\bv{S})-\mu I(\bv{X};\bv{Y}_2|\rv{V},\bv{S})  = h(\bv{X}+\bv{Z}_1|\rv{V},\bv{S})-h(\bv{Z}_1) -\mu\left(h(\bv{X}+\bv{Z}_2|\rv{V},\bv{S})-h(\bv{Z}_2) \right). \nonumber
\end{equation}
Consider the following optimization problem, denoted as $ P $:
\begin{equation} \label{eq:difference_of_entropies}
\sup_{
	\substack{
		P_{\scaleto{\rv{V}\bv{X}|\bv{S}}{4pt}}:
		\rv{V}\rightarrow (\bv{X},\bv{S})\rightarrow (\bv{Y}_1,\bv{Y}_2) \\
		\Exp{\bv{X} \bv{X}^T} \preceq K
	}
} h(\bv{X}+\bv{Z}_1|\rv{V},\bv{S})-\mu h(\bv{X}+\bv{Z}_2|\rv{V},\bv{S}).
\end{equation}
In Section \ref{section:state_dependent_extremal_inequality} we show using a conditional version of an extremal inequality that a conditional Gaussian distribution $ P_{\bv{X}|\rv{V}\bv{S}} \sim \mathcal{N} (0,K_{\bv{X}_1})$ maximizes \eqref{eq:difference_of_entropies}, that is
\begin{equation} \label{eq:maximum_of_entropy_difference}
\sup_{
	\substack{
		P_{\scaleto{\rv{V}\bv{X}|\bv{S}}{4pt}}:
		\rv{V}\rightarrow (\bv{X},\bv{S})\rightarrow (\bv{Y}_1,\bv{Y}_2) \\
		\Exp{\bv{X} \bv{X}^T} \preceq K
	}
}  
h(\bv{X}+\bv{Z}_1|\rv{V},\bv{S})-\mu h(\bv{X}+\bv{Z}_2|\rv{V},\bv{S})
= \frac{1}{2} \log (2\pi e)^t \left| K_{\bv{X}_1} +K_{\bv{Z}_1} \right| -\frac{\mu}{2} \log (2\pi e)^t \left| K_{\bv{X}_1}+K_{\bv{Z}_2} \right|.
\end{equation}
Finally, by collecting \eqref{eq:maximum_entropy_XpZ2_given_S},  \eqref{eq:entropy_Z2}, \eqref{eq:difference_of_mutual_informations} and \eqref{eq:maximum_of_entropy_difference}, we obtain
\begin{align*}
R_1 + \mu R_2 &\leq  \frac{1}{2} \log \frac{\left| K_{\bv{X}_1} +K_{\bv{Z}_1} \right|}{\lvert K_{\bv{Z}_1} \rvert} -\frac{\mu}{2} \log \frac{\left| K_{\bv{X}_1}+K_{\bv{Z}_2} \right|}{\lvert K_{\bv{Z}_2} \rvert} +\frac{\mu}{2} \log \frac{\left| K  - \Sigma_{\bv{X}\bv{S}} K_{\bv{S}}^{-1}\Sigma_{\bv{X}\bv{S}}^T+K_{\bv{Z}_2} \right|}{\left| K_{\bv{Z}_2} \right|} \\
&=  \frac{1}{2} \log \frac{\left| K_{\bv{X}_1} +K_{\bv{Z}_1} \right|}{\lvert K_{\bv{Z}_1} \rvert} +\frac{\mu}{2} \log \frac{\left| K  - \Sigma_{\bv{X}\bv{S}} K_{\bv{S}}^{-1}\Sigma_{\bv{X}\bv{S}}^T+K_{\bv{Z}_2} \right|}{\lvert K_{\bv{X}_1}+K_{\bv{Z}_2} \rvert}.
\end{align*}
Next we proceed to lower bound the leakage rates for $ k \in \{1,2\} $
\begin{equation} \label{eq:mutual_information_state_received_as_entropies}
I(\bv{S};\bv{Y}_k) = h(\bv{S})-h(\bv{S}|\bv{Y}_k).
\end{equation}
The conditional differential entropy can be upper bounded as follows
\begin{align*}
h(\bv{S}|\bv{Y}_k) \leq \frac{1}{2} \log (2\pi e)^2 \left| K_{\bv{S}}-\Sigma_{\bv{S} \bv{Y}_k} \Sigma_{\bv{Y}_k}^{-1} \Sigma_{\bv{S} \bv{Y}_k}^T \right|,
\end{align*}
where $	\Sigma_{\bv{S} \bv{Y}_k} = \Sigma_{\bv{X} \bv{S}}^T +\Sigma_{\bv{S}\bv{S}_k} $ 	and $ \Sigma_{\bv{Y}_k } = K + \Sigma_{\bv{X} \bv{S}_k}+\Sigma_{ \bv{X}\bv{S}_k}^T +K_{\bv{S}_k}+K_{\bv{Z}_k} $.	Hence
\begin{align}
|K_{\bv{S}}-\Sigma_{\bv{S} \bv{Y}_k}\Sigma_{\bv{Y}_k }^{-1} \Sigma_{\bv{S} \bv{Y}_k}^T| 
&=|K_{\bv{S}}||I-K_{\bv{S}}^{-1}\Sigma_{\bv{S} \bv{Y}_k}\Sigma_{\bv{Y}_k }^{-1} \Sigma_{\bv{S} \bv{Y}_k}^T| \nonumber \\
&=|K_{\bv{S}}||I- \Sigma_{\bv{S} \bv{Y}_k}^T K_{\bv{S}}^{-1}\Sigma_{\bv{S} \bv{Y}_k}\Sigma_{\bv{Y}_k }^{-1}| \nonumber \\
&=|K_{\bv{S}}||\Sigma_{\bv{Y}_k }- \Sigma_{\bv{S} \bv{Y}_k}^T K_{\bv{S}}^{-1}\Sigma_{\bv{S} \bv{Y}_k}| |\Sigma_{\bv{Y}_k }^{-1}| \label{eq:mmse_state_given_received},
\end{align}
where the second equality is due to Sylvester's Identity Theorem. The middle determinant in the last equality can be further reformulated as follows
\begin{align}
|\Sigma_{\bv{Y}_k }- \Sigma_{\bv{S} \bv{Y}_k}^T K_{\bv{S}}^{-1}\Sigma_{\bv{S} \bv{Y}_k}| &= |K + \Sigma_{\bv{X} \bv{S}_k}+\Sigma_{\bv{X} \bv{S}_k }^T +K_{\bv{S}_k }+K_{\bv{Z}_k} - ( \Sigma_{\bv{X} \bv{S}}^T +\Sigma_{\bv{S}\bv{S}_k})^T K_{\bv{S}}^{-1}( \Sigma_{\bv{X}\bv{S}}^T +\Sigma_{\bv{S}\bv{S}_k})| \nonumber \\
&= |K- \Sigma_{\bv{X} \bv{S}}  K_{\bv{S}}^{-1} \Sigma_{\bv{X} \bv{S}}^T+K_{\bv{Z}_k} |. \label{eq:mmse_received_given_state}
\end{align} 
Gathering \eqref{eq:mutual_information_state_received_as_entropies}, \eqref{eq:mmse_state_given_received} and \eqref{eq:mmse_received_given_state}, we obtain
\begin{equation} \label{eq:outer_bound_leakage_rate}
I(\bv{S};\bv{Y}_k) \geq \frac{1}{2} \log \frac{|K +\Sigma_{\bv{X} \bv{S}_k}+\Sigma_{\bv{X} \bv{S}_k}^T +K_{\bv{S}_k }+K_{\bv{Z}_k}|} {| K- \Sigma_{\bv{X}\bv{S}} K_{\bv{S}}^{-1} \Sigma_{\bv{X} \bv{S}}^T +K_{\bv{Z}_k} |},
\end{equation}
which concludes the proof of the converse part of Theorem \ref{theorem:achivable_region_mimo_bc_private}.
\subsection{Proof of the direct part of Theorem \ref{theorem:achivable_region_mimo_bc_private}}
Let
\begin{align}
\bv{X} &= \bv{X}_1+\bv{X}_2+B_1 \bv{S}_1+B_2 \bv{S}_2 ,\\
\bv{U} & = \bv{X}_1+A_{10} \bv{X}_2+A_{11} \bv{S}_1+A_{12} \bv{S}_2, \\
\bv{V} & = \bv{X}_2+A_{21} \bv{S}_1+A_{22} \bv{S}_2,
\end{align}
such that $ \bv{X}_1 \sim \mathcal{N} (0,K_{\bv{X}_1}) $, $ \bv{X}_2 \sim \mathcal{N} (0,K_{\bv{X}_2}) $, $ \bv{S}_1 \sim \mathcal{N} (0,K_{\bv{S}_1}) $ and $ \bv{S}_2 \sim \mathcal{N} (0,K_{\bv{S}_2}) $ are mutually independent.

The achievability of Theorem \ref{theorem:achivable_region_mimo_bc_private} follows by evaluating  \eqref{eq:SDDMBC_inner_bound} with the above choice of Gaussian random vectors and the following choice of matrix coefficients
\begin{align}
A_{10} &= K_{\bv{X}_1} ( K_{\bv{X}_1} +K_{\bv{Z}_1})^{-1}, \\
A_{11} &= K_{\bv{X}_1} ( K_{\bv{X}_1} +K_{\bv{Z}_1})^{-1} (B_1+I), \\
A_{12} &= K_{\bv{X}_1} ( K_{\bv{X}_1} +K_{\bv{Z}_1})^{-1} B_2, \\
A_{21} &= K_{\bv{X}_2} ( K_{\bv{X}_1}+K_{\bv{X}_2}+K_{\bv{Z}_2})^{-1} B_1, \\
A_{22} &= K_{\bv{X}_2} ( K_{\bv{X}_1}+K_{\bv{X}_2}+K_{\bv{Z}_2})^{-1}  (B_2+I)\\
B_k &= \Sigma_{\bv{X} \bv{S}_k} \Sigma_{\bv{S}_k}^{-1} \qquad k\in \{1,2\}.
\end{align}
The main idea for this choice of coefficients is to eliminate the state variables from the mutual information terms in Lemma \ref{lemma:inner_bound_dmbc}. The complete proof is given in Appendix \ref{appendix:direct_part_proof}.


\section{An Extremal Inequality} \label{section:state_dependent_extremal_inequality}
In this section we give a sketch of the proof of the conditional extremal inequality, i.e., $ P_{\bv{X}|\rv{V}\bv{S}} \sim \mathcal{N} (0,K_{\bv{X}_1})$ is the solution to $ P $ \eqref{eq:difference_of_entropies}. The idea is a simple extension to the original extremal inequality given in \cite{nair2007outer}. The complementary proof is given in Appendix \ref{section:state_dependent_extremal_inequality_proof}.

Assume $ \bv{Z}_1 \preceq \bv{Z}_2 $. Let $ \bv{Z} $ be such that $ \bv{Z}_2=\bv{Z}_1+\bv{Z} $, where $ \bv{Z} \sim \mathcal{N}(0,K_{\bv{Z}}=K_{\bv{Z}_2}-K_{\bv{Z}_1}) $. The main tool used in the proof is the conditional EPI by Bergmans \cite{bergmans1974}, for which equality in
\begin{equation}
e^{\frac{2}{t}h(\bv{X}+\bv{Z}_1+\bv{Z}|\rv{V},\bv{S})} \geq e^{\frac{2}{t}h(\bv{X}+\bv{Z}_1|\rv{V},\bv{S})}+e^{\frac{2}{t}h(\bv{Z})}
\end{equation}
holds iff $ P_{\bv{X}|\rv{V}\bv{S}} \sim \mathcal{N} (0,K_{\bv{X}_1})$ with the same $ K_{\bv{X}_1} $ for every $ (\rv{V}=v,\bv{S}=\bv{s}) $ and $ K_{\bv{X}_1}+K_{\bv{Z}_1} $ is proportional to $ K_{\bv{Z}} $. The problem is that the proportionality condition is not always satisfied. Hence we introduce the enhanced channel
\begin{align*}
\tilde{\bv{Y}}_1 &= \bv{X}+\bv{S}_1+\tilde{\bv{Z}_1} \\
\tilde{\bv{Y}}_2 &= \bv{X}+\bv{S}_2+\tilde{\bv{Z}_2}.
\end{align*}
where $ \tilde{\bv{Z}}_k \sim \mathcal{N}(0,K_{\tilde{\bv{Z}}_k}) $, $ k\in \{1,2\} $ are constructed such that
\begin{subequations} \label{eq:partial_ordering}
	\begin{align}
	0 \preceq K_{\tilde{\bv{Z}}_1} \preceq K_{\bv{Z}_1},\\
	K_{\tilde{\bv{Z}}_1} \preceq K_{\tilde{\bv{Z}}_2} \preceq K_{\bv{Z}_2},
	\end{align}
\end{subequations} 
and the following proportionality condition is satisfied
\begin{equation*}
K_{\bv{X}_1} +K_{\tilde{\bv{Z}}_1}=(\mu-1)^{-1}K_{\tilde{\bv{Z}}}.
\end{equation*}
Hence, for the enhanced channel, the conditional EPI holds with equality for $ P_{\bv{X}|\rv{V}\bv{S}} \sim \mathcal{N} (0,K_{\bv{X}_1})$. Furthermore, it is straightforward to show that this distribution solves the following equivalent optimization problem $ \tilde{P} $ of \eqref{eq:difference_of_mutual_informations} applied to the enhanced channel, i.e.,
\begin{equation*}
\max_{P_{\bv{X}|\rv{V},\bv{S}}:\Exp[]{\rv{X}^2} \preceq K} h(\bv{X}+\tilde{\bv{Z}}_1|\rv{V},\bv{S})-\mu h(\bv{X}+\tilde{\bv{Z}}_2|\rv{V},\bv{S})+F
\end{equation*}
where the constant $ F $ is introduced to make the optimum value of the objective functions of $ P_G $, which is $ P $ constrained to conditional Gaussian $ P_{\bv{X}|\rv{V}\bv{S}} $, and $ \tilde{P} $ coincide.

It remains to show, that the objective function of $ P $ is less than or equal to the objective function of $ \tilde{P} $ for any choice of $ P_{\bv{X}|\rv{V}\bv{S}} $. This observation follows from the partial ordering in \eqref{eq:partial_ordering} and application of Data Processing Inequality and Worst Additive Noise Lemma \cite[Lemma II.2]{diggavi2001worst}.

To conclude, we have shown that $ (P_G)=(\tilde{P}) $. Furthermore $ P_G \leq P \leq \tilde{P} $ for any choice of $ P_{\bv{X}|\rv{V}\bv{S}} $. Thus, $ P_{\bv{X}|\rv{V}\bv{S}} \sim \mathcal{N}(0,K_{\bv{X}_1}) $ also solves $ P $.


\section{Conclusions}
In this paper we addressed the problem of simultaneous communication and state masking over a MIMO Gaussian BC with additive interference modeled as state and given as a noncausal CSI to the encoder. We developed a new conditional extremal inequality in order to characterize the achievable region for the private messages scenario. Moreover, the standard results of point-to-point masking \cite{merhav2007information} and state-dependent BC \cite{steinberg2005achievable} (no masking requirements), emerge as special cases of the bounds here. An extension to the MIMO Gaussian BC with an additional common message is under current study.


\section*{Acknowledgment}
The work of M. Dikshtein and S. Shamai (Shitz) has been supported by the
European Union's Horizon 2020 Research And Innovation Programme,
grant agreement no. 694630.
The work of A. Somekh-Baruch and S. Shamai (Shitz) was also
supported by the Heron consortium via the Israel ministary of economy
and science.

\appendices

\section{} \label{appendix:direct_part_proof}
Let $ \bv{S} \triangleq (\bv{S}_1,\bv{S}_2) $ and let
\begin{align}
\bv{X} &= \bv{X}_1+\bv{X}_2+B_1 \bv{S}_1+B_2 \bv{S}_2 ,\\
\bv{U} & = \bv{X}_1+A_{10} \bv{X}_2+A_{11} \bv{S}_1+A_{12} \bv{S}_2, \\
\bv{V} & = \bv{X}_2+A_{21} \bv{S}_1+A_{22} \bv{S}_2,
\end{align}
such that $ \bv{X}_1 \sim \mathcal{N} (0,K_1) $, $ \bv{X}_2 \sim \mathcal{N} (0,K_2) $, $ \bv{S}_1 \sim \mathcal{N} (0,\Sigma_{\bv{S}_1}) $ and $ \bv{S}_2 \sim \mathcal{N} (0,\Sigma_{\bv{S}_2}) $ are mutually independent. We also define
\begin{equation}
M_{\bv{X}|\bv{Y}} \triangleq \Exp[]{\bv{X} \bv{Y}^T} (\Exp{\bv{Y}\bv{Y}^T})^{-1}.
\end{equation}

With these definitions and applying Lemma \ref{lemma:inner_bound_dmbc}, the achievability of $ R_2 $ in the RHS of \eqref{eq:mimo_bc_r2}  can be shown as follows:
\begin{equation}
I(\bv{V};\bv{Y}_2)-I(\bv{V};\bv{S}) = h(\bv{V}|\bv{S})-h(\bv{V}|\bv{Y}_2),
\end{equation}
where $ h(\bv{V}|\bv{S}) = h(\bv{X}_2) $ and
\begin{align}
h(\bv{V}|\bv{Y}_2) &= h(\bv{V}-\Exp{\bv{V}|\bv{Y}_2}|\bv{Y}_2) \nonumber \\
&\eqann[=]{a} h(\bv{V}-\Exp{\bv{V}|\bv{Y}_2}) \nonumber \\
&\eqann[=]{b} h(\bv{V}- M_{\bv{V} | \bv{Y}_2} \bv{Y}_2).
\end{align}
where \eqannref{a} and \eqannref{b} follow since $ (\rv{V},\bv{Y}_2) $ are jointly Gaussian.

In order to obtain the WDP property, we require that $ \bv{V}-M_{\bv{V} | \bv{Y}_2} \bv{Y}_2 $ would not contain $ \bv{S}_1 $ or $ \bv{S_2} $, hence
\begin{equation}
\bv{V}-M_{\bv{V} | \bv{Y}_2} \bv{Y}_2 = \bv{X}_2+A_{21} \bv{S}_1+A_{22} \bv{S}_2 -M_{\bv{V} | \bv{Y}_2} \big( \bv{X}_1+\bv{X}_2+B_1 \bv{S}_1+B_2 \bv{S}_2+\bv{S}_2+\bv{Z}_2 \big). \nonumber
\end{equation}
Thus
\begin{align}
A_{21} &= M_{\bv{V} | \bv{Y}_2} B_1, \\
A_{22} &= M_{\bv{V} | \bv{Y}_2}  (B_2+I),
\end{align}
with such choice of $ A_{21} $ and $ A_{22} $ we have
\begin{equation}
\bv{V}-M_{\bv{V} | \bv{Y}_2} \bv{Y}_2 = \bv{X}_2 -M_{\bv{V} | \bv{Y}_2} \big( \bv{X}_1+\bv{X}_2 +\bv{Z}_2 \big).
\end{equation}
We proceed by requiring that $ M_{\bv{V} | \bv{Y}_2} $ would be the MMSE estimator of $ \bv{X}_2 $ given $ \bv{X}_1+\bv{X_2}+\bv{Z}_2 $,
\begin{equation}
M_{\bv{V} | \bv{Y}_2} = K_{\bv{X}_2} (  K_{\bv{X}_1}+K_{\bv{X}_2} +K_{\bv{Z}_2})^{-1}.
\end{equation}
Thus
\begin{align}
A_{21} &=K_{\bv{X}_2} ( K_{\bv{X}_1}+K_{\bv{X}_2}+K_{\bv{Z}_2})^{-1} B_1, \\
A_{22} &=K_{\bv{X}_2} ( K_{\bv{X}_1}+K_{\bv{X}_2}+K_{\bv{Z}_2})^{-1}  (B_2+I)
\end{align}
and
\begin{equation}
h(\bv{V}|\bv{Y}_2) = h(\bv{X}_2|\bv{X_1+\bv{X}_2}+\bv{Z}_2).
\end{equation}
Finally
\begin{align}
I(\bv{V};\bv{Y}_2)-I(\bv{V};\bv{S}) &= h(\bv{X}_2)-h(\bv{X}_2|\bv{X_1+\bv{X}_2}+\bv{Z}_2) \nonumber \\
&= I(\bv{X}_2;\bv{X_1+\bv{X}_2}+\bv{Z}_2) \nonumber \\
&= h(\bv{X_1+\bv{X}_2}+\bv{Z}_2)-h(\bv{X}_1+\bv{Z}_2) \nonumber \\
&= \frac{1}{2}\log \frac{|K_{\bv{X}_1}+K_{\bv{X}_2}+K_{\bv{Z}_2}|}{|K_{\bv{X}_1}+K_{\bv{Z}_2}|}.
\end{align}
Similarly, the RHS of \eqref{eq:mimo_bc_r1} can be achieved as follows:
\begin{equation}
I(\bv{U};\bv{Y}_1)-I(\bv{U};\bv{V},\bv{S}) = h(\bv{U}|\bv{V},\bv{S})-h(\bv{U}|\bv{Y}_1),
\end{equation}
where $ h(\bv{U}|\bv{V},\bv{S}) = h(\bv{X}_1) $ and
\begin{align}
h(\bv{U}|\bv{Y}_1) &= h(\bv{U}-\Exp{\bv{U}|\bv{Y}_1}|\bv{Y}_1) \nonumber \\
&\eqann[=]{a} h(\bv{U}-\Exp{\bv{U}|\bv{Y}_1}) \nonumber \\
&\eqann[=]{b} h(\bv{U}- M_{\rv{U}|\bv{Y}_1}\bv{Y}_1).
\end{align}
where \eqannref{a} and \eqannref{b} follow since $ (\rv{U},\bv{Y}_1) $ are jointly Gaussian.

Similarly as for $ R_2 $, we require that $ \bv{U}-M_{\rv{U}|\bv{Y}_1}\bv{Y}_1 $ would not contain $ \bv{X}_2 $, $ \bv{S}_1 $ or $\bv{S}_2 $, hence
\begin{equation*}
\bv{U}-M_{\rv{U}|\bv{Y}_1} \bv{Y}_1 = \bv{X}_1+A_{10}\bv{X}_2+A_{11} \bv{S}_1+A_{12} \bv{S}_2 -M_{\rv{U}|\bv{Y}_1} \big( \bv{X}_1+\bv{X}_2+B_1 \bv{S}_1+B_2 \bv{S}_2+\bv{S}_1+\bv{Z}_1 \big).
\end{equation*}
Thus
\begin{align}
A_{10} &=M_{\rv{U}|\bv{Y}_1} , \\
A_{11} &=M_{\rv{U}|\bv{Y}_1} (B_1+I), \\
A_{12} &=M_{\rv{U}|\bv{Y}_1}   B_2,
\end{align}
with such choice of $ A_{10} $, $ A_{11} $ and $ A_{12} $ we have
\begin{equation}
\rv{U}-M_{\rv{U}|\bv{Y}_1} \bv{Y}_1 = \bv{X}_1 -M_{\rv{U}|\bv{Y}_1} \big(  \bv{X}_1+\bv{Z}_1 \big).
\end{equation}
We proceed by requiring that $ M_{\rv{U}|\bv{Y}_1} $ would be the MMSE estimator of $ \bv{X}_1 $ given $ \bv{X_1}+\bv{Z}_1 $, hence
\begin{equation}
M_{\rv{U}|\bv{Y}_1} = K_{\bv{X}_1} ( K_{\bv{X}_1} +K_{\bv{Z}_1})^{-1}.
\end{equation}
Thus
\begin{align}
A_{10} &= K_{\bv{X}_1} ( K_{\bv{X}_1} +K_{\bv{Z}_1})^{-1}, \\
A_{11} &= K_{\bv{X}_1} ( K_{\bv{X}_1} +K_{\bv{Z}_1})^{-1} (B_1+I), \\
A_{12} &= K_{\bv{X}_1} ( K_{\bv{X}_1} +K_{\bv{Z}_1})^{-1} B_2,
\end{align}
and
\begin{equation} \label{eq:mimo_common_centropy_u_given_y1}
h(\rv{U}|\bv{Y}_1) = h(\bv{X}_1| \bv{X}_1+\bv{Z}_1).
\end{equation}
Hence
\begin{align}
I(\rv{U};\bv{Y}_1)-I(\rv{U};\rv{V},\bv{S}) &= h(\bv{X}_1)-h(\bv{X}_1|\bv{X}_1+\bv{Z}_1) \nonumber \\
&= I(\bv{X}_1;\bv{X}_1+\bv{Z}_1) \nonumber \\
&= h( \bv{X}_1+\bv{Z}_1)-h(\bv{Z}_1) \nonumber \\
&= \frac{1}{2}\log \frac{| K_{\bv{X}_1} +K_{\bv{Z}_1}|}{|K_{\bv{Z}_1}|},
\end{align}

We proceed to show achievability of $ E_2 $ in the RHS of \eqref{eq:mimo_bc_e2}. Consider the achievable masking rate in \eqref{eq:SDDMBC_inner_bound},
\begin{equation}
I(\bv{S};\rv{V},\bv{Y}_2) =I(\bv{S};\bv{Y}_2)+I(\bv{S};\rv{V}|\bv{Y}_2).
\end{equation}
First we show that the second mutual information term is zero
\begin{align*}
I(\bv{S};\rv{V}|\bv{Y}_2) &= h(\bv{S}|\bv{Y}_2)-h(\bv{S}|\rv{V},\bv{Y}_2) \\
&= h(\bv{S}|\bv{Y}_2)-h(\bv{S}|\rv{V}-M_{\rv{V}|\bv{Y}_2}\bv{Y}_2,\bv{Y}_2) \\
&= h(\bv{S}|\bv{Y}_2)-h(\bv{S}|\bv{X}_2-M_{\rv{V}|\bv{Y}_2}(\bv{X}_1+\bv{X}_2+\bv{Z}_2),\bv{Y}_2) .
\end{align*}
Denote $ \tilde{\bv{X}}_2 \triangleq \bv{X}_2-M_{\rv{V}|\bv{Y}_2}(\bv{X}_1+\bv{X}_2+\bv{Z}_2) $,
\begin{equation*}
h(\bv{S}|\tilde{\bv{X}}_2,\bv{Y}_2)
=h \left(\bv{S} -
\begin{pmatrix}
\Sigma_{\bv{S} \tilde{\bv{X}}_2} & \Sigma_{\bv{S} \bv{Y}_2}
\end{pmatrix}
\begin{pmatrix}
\Sigma_{\tilde{\bv{X}}_2} & \Sigma_{\tilde{\bv{X}}_2 \bv{Y}_2} \\
\Sigma_{\tilde{\bv{X}}_2 \bv{Y}_2} & \Sigma_{\bv{Y}_2}
\end{pmatrix}^{-1}
\begin{pmatrix}
\tilde{\bv{\bv{X}}}_2 \\
\bv{Y}_2
\end{pmatrix}
\right).
\end{equation*}
It is straightforward to show that $ \Sigma_{\bv{S} \tilde{\bv{X}}_2}=0 $ and $ \Sigma_{\tilde{\bv{X}}_2 \bv{Y}_2}=0 $, hence
\begin{align*}
h(\bv{S}|\tilde{\bv{X}}_2,\bv{Y}_2) &=h \left(\bv{S} -
\begin{pmatrix}
0 \\
\Sigma_{\bv{S} \bv{Y}_2} \Sigma_{\bv{Y}_2}^{-1}
\end{pmatrix}
\begin{pmatrix}
\tilde{\bv{X}}_2 \\
\bv{Y}_2
\end{pmatrix}
\right) \\
&=h \left(\bv{S} -\Sigma_{\bv{S} \bv{Y}_2} \Sigma_{\bv{Y}_2}^{-1} \bv{Y}_2		\right) \\
&=h \left(\bv{S} | \bv{Y}_2		\right).
\end{align*}
Hence
\begin{equation}
I(\bv{S};\rv{V}|\bv{Y}_2)=0.
\end{equation}
Now we evaluate $ I(\bv{S};\bv{Y}_2) $
\begin{align*}
I(\bv{S};\bv{Y}_2) &= h(\bv{Y}_2)-h(\bv{Y}_2|\bv{S}) \\
&= h(\bv{Y}_2) -h(\bv{X}_1+\bv{X}_2+\bv{Z}_2) \\
&= \frac{1}{2} \log \frac{|K +(B_2+I) K_{\bv{S}_2 }+K_{\bv{S}_2}B_2^T  +K_{\bv{Z}_2}|}{|K_{\bv{X}_1}+K_{\bv{X}_2}+K_{\bv{Z}_2}|}.
\end{align*}
In order to have similar expression as in the converse part, we choose
\begin{equation}
B_k = \Sigma_{\bv{X} \bv{S}_k} K_{\bv{S}_k}^{-1} \qquad k\in \{1,2\}.
\end{equation}
Thus,
\begin{align*}
K &= K_{\bv{X}_1}+K_{\bv{X}_2}+\Sigma_{\bv{X} \bv{S}_1} K_{\bv{S}_1}^{-1} \Sigma_{\bv{X} \bv{S}_1}^T+\Sigma_{\bv{X} \bv{S}_2} K_{\bv{S}_2}^{-1} \Sigma_{\bv{X} \bv{S}_2}^T\\
&= K_{\bv{X}_1}+K_{\bv{X}_2}+\Sigma_{\bv{X} \bv{S}} K_{\bv{S}}^{-1} \Sigma_{\bv{X} \bv{S}}^T
\end{align*}
where $ K_{\bv{S}} $ and $ \Sigma_{ \bv{X}\bv{S}} $ were defined in \eqref{eq:K_S_definition} and \eqref{eq:Sigma_X_S_definition} respectively.
Finally
\begin{equation}
I(\bv{S};\rv{V},\bv{Y}_2) = \frac{1}{2} \log \frac{|K +\Sigma_{\bv{X} \bv{S}_2}+\Sigma_{\bv{X} \bv{S}_2 }^T +K_{\bv{S}_2}+K_{\bv{Z}_2}|} {|K- \Sigma_{\bv{X}\bv{S}}  K_{\bv{S}}^{-1} \Sigma_{\bv{X}\bv{S}}^T +K_{\bv{Z}_2} |}, \nonumber
\end{equation}
which is identical to the outer bound expression \eqref{eq:outer_bound_leakage_rate}.
Similarly, it can be shown that
\begin{equation}
I(\bv{S};\rv{U},\bv{Y}_1) = \frac{1}{2} \log \frac{\lvert K +\Sigma_{\bv{X} \bv{S}_1}+\Sigma_{\bv{X} \bv{S}_1 }^T +K_{\bv{S}_1}+K_{\bv{Z}_1} \rvert} {|K- \Sigma_{\bv{X}\bv{S}}  K_{\bv{S}}^{-1} \Sigma_{\bv{X}\bv{S}}^T +K_{\bv{Z}_1} |}. \nonumber
\end{equation}

\section{} \label{section:state_dependent_extremal_inequality_proof}
We would like to show that conditional Gaussian density is the maximizing density of $ P $ \eqref{eq:difference_of_entropies},
\begin{equation} \label{eq:op_P}
\max_{P_{\bv{X}|\rv{V},\bv{S}}:\Exp[]{\bv{X} \bv{X}^T} \preceq K} h(\bv{X}+\bv{Z}_1|\rv{V},\bv{S})-\mu h(\bv{X}+\bv{Z}_2|\rv{V},\bv{S}),
\end{equation}
where $ \bv{S} \sim \mathcal{N}(0,K_{\bv{S}}) $, $ \bv{Z}_1 \sim \mathcal{N}(0,K_{\bv{Z}_1}) $ and $ \bv{Z}_2 \sim \mathcal{N}(0,K_{\bv{Z}_2}) $. $ K_{\bv{Z}_2} \succeq K_{\bv{Z}_1} $, hence we can write $ \bv{Z}_2=\bv{Z}_1+\bv{Z} $, where $ \bv{Z}\sim \mathcal{N}(0,K_{\bv{Z}_2}-K_{\bv{Z}_1}) $. We denote the optimal value of $ P $ by $ (P) $. We utilize the following conditional version of the EPI \cite{bergmans1974}
\begin{align*}
h(\bv{X}+\bv{Z}_2|\rv{V},\bv{S}) &= h(\bv{X}+\bv{Z}_1+\bv{Z}|\rv{V},\bv{S}) \\
&\geq \frac{t}{2} \log \left( e^{\frac{2}{t}h(\bv{X}+\bv{Z}_1|\rv{V},\bv{S})} + e^{\frac{2}{t}h(\bv{Z})} \right),
\end{align*}
where equality holds if and only if conditioning on $ (\rv{V}=v,\bv{S}=\bv{s}) $, $ \bv{X}+\bv{Z}_1 $ is Gaussian with the same covariance matrix for each $ (v,\bv{s}) $ and proportional to that of $ \bv{Z} $, that is, there exists $ \gamma\in \mathbb{R} $, such that
\begin{equation} \label{eq:proportionality_condition}
K_{\bv{X}}+K_{\bv{Z}_1} = \gamma K_{\bv{Z}}.
\end{equation} 
We also denote by $ P_G $ the Gaussian version of $ P $ by restricting the solution space to Gaussian distribution, i.e.
\begin{equation*}
\max_{K_{\bv{X}} \preceq K} \frac{1}{2} \log \left(  (2\pi e)^t \lvert K_{\bv{X}}+K_{\bv{Z}_1}\rvert \right) - \frac{\mu}{2} \log \left(  (2\pi e)^t \lvert K_{\bv{X}}+K_{\bv{Z}_2}\rvert \right).
\end{equation*}
The optimal solution of $ P_G $, $ K_{\bv{X}}^{\star} $, must satisfy the following KKT-like conditions
\begin{subequations} \label{eq:KKT}
	\begin{equation}
	\frac{1}{2} \left(K_{\bv{X}}^{\star} +K_{\bv{Z}_1}\right)^{-1}+M_1=\frac{\mu}{2} \left(K_{\bv{X}}^{\star} +K_{\bv{Z}_2}\right)^{-1}+M_2 ,\label{eq:KKT1}
	\end{equation}
	\begin{equation}
	M_1 K_{\bv{X}}^{\star} = 0 ,\label{eq:KKT2}
	\end{equation}
	\begin{equation}
	M_2(K-K_{\bv{X}}^{\star}) = 0. \label{eq:KKT3}
	\end{equation}
\end{subequations}
The channel parameters do not necessarily comply with the proportionality condition.
Let $ K_{\tilde{\bv{Z}}_1} $ and $ K_{\tilde{\bv{Z}}_2} $, be two real symmetric matrices satisfying
\begin{subequations} \label{eq:enhancement_conditions}
	\begin{equation} \label{eq:enhancement_condition_Z1}
	\frac{1}{2} \left(K_{\bv{X}}^{\star} +K_{\bv{Z}_1}\right)^{-1}+M_1 =\frac{1}{2} \left(K_{\bv{X}}^{\star} +K_{\tilde{\bv{Z}}_1}\right)^{-1} ,
	\end{equation}
	\begin{equation} \label{eq:enhancement_condition_Z2}
	\frac{\mu}{2} \left(K_{\bv{X}}^{\star} +K_{\bv{Z}_2}\right)^{-1}+M_2 = \frac{\mu}{2} \left(K_{\bv{X}}^{\star} +K_{\tilde{\bv{Z}}_2}\right)^{-1}.
	\end{equation}
\end{subequations}
We define the following auxiliary optimization problem $ \tilde{P} $ with optimum value $ (\tilde{P}) $ for which we show next that the condition in \eqref{eq:proportionality_condition} holds
\begin{equation} \label{eq:enhanced_channel_optimization_problem}
\max_{P_{\bv{X}|\rv{V},\bv{S}}:\Exp[]{\bv{X} \bv{X}^T} \preceq K} h(\bv{X}+\tilde{\bv{Z}}_1|\rv{V},\bv{S})-\mu h(\bv{X}+\tilde{\bv{Z}}_2|\rv{V},\bv{S})+F.
\end{equation}
The constant $ F $ is defined as
\begin{align*}
F \triangleq h(\bv{Z}_1)-h(\tilde{\bv{Z}}_1)+\mu (h(\bv{X}_{G_K}+\tilde{\bv{Z}}_2)-h(\bv{X}_{G_K}+\bv{Z}_2)),
\end{align*}
and $ \bv{X}_{G_K} \sim \mathcal{N}(0,K) $, $ \tilde{\bv{Z}}_1 \sim \mathcal{N}(0,K_{\tilde{\bv{Z}}_1}) $ and $ \tilde{\bv{Z}}_2 \sim \mathcal{N}(0,K_{\tilde{\bv{Z}}_1}) $. $ K_{\tilde{\bv{Z}}_2} \succeq K_{\tilde{\bv{Z}}_1} $.

It was shown in \cite{liu2007extremal} that $ 0 \preceq K_{\tilde{\bv{Z}}_1} \preceq K_{\bv{Z}_1} $ and $ K_{\tilde{\bv{Z}}_1} \preceq K_{\tilde{\bv{Z}}_2} \preceq K_{\bv{Z}_2} $, hence we can write $ \tilde{\bv{Z}}_2=\tilde{\bv{Z}}_1+\tilde{\bv{Z}} $, where $ \tilde{\bv{Z}}\sim \mathcal{N}(0,K_{\tilde{\bv{Z}}}=K_{\tilde{\bv{Z}}_2}-K_{\tilde{\bv{Z}}_1}) $. By substituting \eqref{eq:enhancement_conditions} into the KKT-like condition \eqref{eq:KKT1} we have
\begin{align*}
\mu \left(K_{\bv{X}}^{\star} +K_{\tilde{\bv{Z}}_1}\right) = K_{\bv{X}}^{\star} +K_{\tilde{\bv{Z}}_2},
\end{align*}
which is equivalent to
\begin{equation} \label{eq:proportionality_condition_satisfied}
K_{\bv{X}}^{\star} +K_{\tilde{\bv{Z}}_1}=(\mu-1)^{-1}K_{\tilde{\bv{Z}}},
\end{equation}
and hence the proportionality condition is satisfied for the enhanced channel.

In what follows we prove that $ P_{\bv{X}|\rv{U}\bv{S}} \sim \mathcal{N}(0,K_{\bv{X}_1}) $ is the solution to $ P $. We will show this in three steps. First we prove that the objective function of $ P $ is less or equal of the objective function of $ \tilde{P} $ for any choice of $ P_{\bv{X}|\rv{V}\bv{S}} $. Then we demonstrate that $ P_{\bv{X}|\rv{U}\bv{S}} \sim \mathcal{N}(0,K_{\bv{X}_1}) $ is the solution of $ \tilde{P} $. Finally we will show that $ P_{\bv{X}|\rv{U}\bv{S}} \sim \mathcal{N}(0,K_{\bv{X}_1}) $ is also the solution of $ P_G $. Since $ P_G \leq P \leq \tilde{P} $, the conditional extremal inequality is established.
\subsection{Step I: $ P<\tilde{P} $}
Here we show that the argument of $ P $ less than or equal to the argument of $ \tilde{P} $ for every $ P_{\bv{X}|\rv{U}\bv{S}} $. The difference between the objective functions of $ P $ \eqref{eq:difference_of_entropies} and $ \tilde{P} $ \eqref{eq:enhanced_channel_optimization_problem} can be written as
\begin{equation*}
h(\bv{X}+\bv{Z}_1|\rv{V},\bv{S})-h(\bv{Z}_1)- h(\bv{X}+\tilde{\bv{Z}}_1|\rv{V},\bv{S})+h(\tilde{\bv{Z}}_1) 
-\mu \left( h(\bv{X}+\bv{Z}_2|\rv{V},\bv{S})-h(\bv{X}_{G_K}+\bv{Z}_2) - h(\bv{X}+\tilde{\bv{Z}}_2|\rv{V},\bv{S})+h(\bv{X}_{G_K}+\tilde{\bv{Z}}_2) \right).
\end{equation*}
As we have already mentioned, from the construction of the enhanced channel, $ K_{\tilde{\bv{Z}}_1} \preceq K_{\bv{Z}_1} $. Let $ \hat{\bv{Z}}_1 \triangleq \bv{Z}_1-\tilde{\bv{Z}}_1 $. We have
\begin{align*}
h(\bv{X}+\bv{Z}_1|\rv{V},\bv{S})-h(\bv{Z}_1)- h(\bv{X}+\tilde{\bv{Z}}_1|\rv{V},\bv{S})+h(\tilde{\bv{Z}}_1) 
&= I(\bv{X};\bv{X}+\bv{Z}_1|\rv{V},\bv{S}) - I(\bv{X};\bv{X}+\tilde{\bv{Z}}_1|\rv{V},\bv{S}) \\
&= I(\bv{X};\bv{X}+\tilde{\bv{Z}}_1+\hat{\bv{Z}}_1|\rv{V},\bv{S}) - I(\bv{X};\bv{X}+\tilde{\bv{Z}}_1|\rv{V},\bv{S}) \\
&\leq 0,
\end{align*}
The last inequality is since given $ (\rv{V},\bv{S}) $, we have the Markov chain 
\begin{equation*}
\bv{X} \rightarrow \bv{X}+\tilde{\bv{Z}}_1 \rightarrow \bv{X}+\tilde{\bv{Z}}_1+\hat{\bv{Z}}_1,
\end{equation*}
and applying Data Processing Inequality \cite[p. 24]{gamal2011network}. It holds for any choice of $ P_{\bv{X}|\rv{V},\bv{S}} $. Similarly $ K_{\tilde{\bv{Z}}_2} \preceq K_{\bv{Z}_2} $. Let $ \hat{\bv{Z}}_2 \triangleq \bv{Z}_2-\tilde{\bv{Z}}_2 $. Further
\begin{align*}
&h(\bv{X}+\bv{Z}_2|\rv{V},\bv{S})- h(\bv{X}+\tilde{\bv{Z}}_2|\rv{V},\bv{S})-h(\bv{X}_{G_K}+\bv{Z}_2) +h(\bv{X}_{G_K}+\tilde{\bv{Z}}_2) \\
&=h(\bv{X}+\tilde{\bv{Z}}_2+\hat{\bv{Z}}_2|\rv{V},\bv{S})- h(\bv{X}+\tilde{\bv{Z}}_2|\rv{V},\bv{S})-h(\bv{X}_{G_K}+\tilde{\bv{Z}}_2+\hat{\bv{Z}}_2) +h(\bv{X}_{G_K}+\tilde{\bv{Z}}_2) \\ 
&=I(\hat{\bv{Z}}_2;\bv{X}+\tilde{\bv{Z}}_2+\hat{\bv{Z}}_2|\rv{V},\bv{S})-I(\hat{\bv{Z}}_2;\bv{X}_{G_K}+\tilde{\bv{Z}}_2+\hat{\bv{Z}}_2) \\
&\eqann[\geq]{a} I(\hat{\bv{Z}}_2;\bv{X}_G+\tilde{\bv{Z}}_2+\hat{\bv{Z}}_2|\rv{V},\bv{S})-I(\hat{\bv{Z}}_2;\bv{X}_{G_K}+\tilde{\bv{Z}}_2+\hat{\bv{Z}}_2) \\
&\eqann[=]{b} I(\hat{\bv{Z}}_2;\bv{X}_G+\tilde{\bv{Z}}_2+\hat{\bv{Z}}_2|\rv{V},\bv{S})-I(\hat{\bv{Z}}_2;\bv{X}_G+\hat{\bv{X}}_G+\tilde{\bv{Z}}_2+\hat{\bv{Z}}_2) \\
&\eqann[\geq]{c} 0,
\end{align*}
where inequality in \eqannref{a} follows from Worst Additive Noise Lemma \cite[Lemma II.2]{diggavi2001worst}, \eqannref{b} is due to $ \bv{X}_{G_K}=\bv{X}_G+\hat{\bv{X}}_G $, $ \hat{\bv{X}}_G $ is independent of $ \bv{X}_G $. Inequality in \eqannref{c} is again due to the Markov chain 
\begin{equation*}
\hat{\bv{Z}}_2 \rightarrow \bv{X}_G+\tilde{\bv{Z}}_2+\hat{\bv{Z}}_2 \rightarrow \bv{X}_G+\hat{\bv{X}}_G+\tilde{\bv{Z}}_2+\hat{\bv{Z}}_2,
\end{equation*}
and Data Processing Inequality \cite[p. 24]{gamal2011network}. Thus, we have shown that the objective function of $ P $ is less or equal to the objective function of $ \tilde{P} $ for any choice of $ P_{\bv{X}|\rv{V}\bv{S}} $.
\subsection{Step II: $ (\tilde{P}) = (\tilde{P}_G) $ }
In this subsection we show that $ P_{\bv{X}|\rv{V}\bv{S}}  \sim \mathcal{N}(0,K_{\bv{X}_1})$ is the solution of $ \tilde{P} $. We begin with the conditional EPI \cite{bergmans1974}, that states
\begin{align*} 
h(\bv{X}+\tilde{\bv{Z}}_2|\rv{V},\bv{S}) &= h(\bv{X}+\tilde{\bv{Z}}_1+\tilde{\bv{Z}}|\rv{V},\bv{S}) \\
&\geq \frac{t}{2} \log \left( e^{\frac{2}{t} h(\bv{X}+\tilde{\bv{Z}}_1|\rv{V},\bv{S})}+e^{\frac{2}{t}h(\tilde{\bv{Z}})} \right).
\end{align*}
Hence, the difference between the conditional differential entropies in \eqref{eq:enhanced_channel_optimization_problem} can be upper bounded as
\begin{equation} \label{eq:entropy_difference_ub1}
h(\bv{X}+\tilde{\bv{Z}}_1|\rv{V},\bv{S})-\mu h(\bv{X}+\tilde{\bv{Z}}_2|\rv{V},\bv{S}) 
\leq h(\bv{X}+\tilde{\bv{Z}}_1|\rv{V},\bv{S})-\frac{\mu t}{2} \log \left( e^{\frac{2}{t} h(\bv{X}+\tilde{\bv{Z}}_1|\rv{V},\bv{S})}+e^{\frac{2}{t}h(\tilde{\bv{Z}})} \right).
\end{equation}
Next we utilize the following function which was defined in \cite{liu2007extremal}
\begin{equation}
f(a,b) = a-\frac{\mu t}{2} \log \left(e^{\frac{2}{t}a}+e^{\frac{2}{t}b}\right).
\end{equation}
It was shown to have a global maximum at $ a=b-\frac{t}{2} \log (\mu-1) $, hence the inequality in \eqref{eq:entropy_difference_ub1} can be rewritten as
\begin{align*}
h(\bv{X}+\tilde{\bv{Z}}_1|\rv{V},\bv{S})-\mu h(\bv{X}+\tilde{\bv{Z}}_2|\rv{V},\bv{S}) 
&\eqann[\leq]{a} f \left(h(\bv{X}+\tilde{\bv{Z}}_1|\rv{V},\bv{S}),h(\tilde{\bv{Z}}) \right) \\
&\eqann[\leq]{b} f \left(h(\tilde{\bv{Z}})-\frac{t}{2} \log (\mu-1),h(\tilde{\bv{Z}}) \right).
\end{align*}
Since $ K_{\bv{X}}^{\star} +K_{\tilde{\bv{Z}}_1}=(\mu-1)^{-1}K_{\tilde{\bv{Z}}} $, equality in \eqannref{a} holds if $ P_{\bv{X}|\rv{V}\bv{S}} \sim \mathcal{N}(0,K_{\bv{X}_1}) $ with same $ K_{\bv{X}_1} $ for all $ (\rv{V}=v,\bv{S}=\bv{s}) $. Equality in $ \eqannref{b} $ holds if $ h(\bv{X}_G+\tilde{\bv{Z}}_1|\rv{V},\bv{S})= h(\tilde{\bv{Z}})-\frac{t}{2} \log (\mu-1) $, i.e.
\begin{equation*}
\frac{1}{2} \log \left((2\pi e)^t \left| K_{\bv{X}_1}+K_{\tilde{\bv{Z}}_1} \right| \right) 
= \frac{1}{2} \log \left((2\pi e)^t \left| K_{\tilde{\bv{Z}}} \right| \right)-\frac{t}{2} \log(\mu-1),
\end{equation*}
which is equivalent to
\begin{align*}
\left| K_{\bv{X}_1}+K_{\tilde{\bv{Z}}_1} \right| = (\mu-1)^{-1} \left| K_{\tilde{\bv{Z}}} \right| .
\end{align*}
The last condition is already satisfied by \eqref{eq:proportionality_condition_satisfied}. Hence $ P_{\bv{X}|\rv{V}\bv{S}} \sim \mathcal{N}(0,K_{\bv{X}_1}) $ with same $ K_{\bv{X}_1} $ for all $ (\rv{V}=v,\bv{S}=\bv{s}) $ is the maximizing distribution of $ \tilde{P} $.

\subsection{Step III: $ (P_G) =  (\tilde{P}) $}
It remains to show that the objective functions of $ P_G $ and $ \tilde{P}_G $ take equal values at $ P_{\bv{X}|\rv{V},\bv{S}} \sim \mathcal{N}(0,K_{\bv{X}_1})$. Let $ \bv{X}_{G_1} $ be denoted as the random vector distributed according to $ \mathcal{N}(0,K_{\bv{X}_1}) $ given $ (\rv{V}=v,\bv{S}=\bv{s}) $. The objective value of $ P_G $ would be then
\begin{equation*}
h(\bv{X}_{G_1}+\bv{Z}_1|\rv{V},\bv{S}) - \mu h(\bv{X}_{G_1}+\bv{Z}_2|\rv{V},\bv{S}) = \frac{1}{2} \log \left((2\pi e)^t\lvert K_{\bv{X}_1}+K_{\bv{Z}_1} \rvert \right) - \frac{\mu}{2} \log \left((2\pi e)^t\lvert K_{\bv{X}_1}+K_{\bv{Z}_2} \rvert \right).
\end{equation*}
The objective value of $ \tilde{P} $ would be then
\begin{align*}
&h(\bv{X}_{G_1}+\tilde{\bv{Z}}_1|\rv{V},\bv{S}) - \mu h(\bv{X}_{G_1}+\tilde{\bv{Z}}_2|\rv{V},\bv{S})+F\\
&= \frac{1}{2} \log \left((2\pi e)^t\lvert K_{\bv{X}_1}+K_{\tilde{\bv{Z}}_1} \rvert \right) - \frac{\mu}{2} \log \left((2\pi e)^t\lvert K_{\bv{X}_1}+K_{\tilde{\bv{Z}}_2} \rvert \right)+\frac{1}{2} \log \frac{|K_{\bv{Z}_1}|}{|K_{\tilde{\bv{Z}}_1}|}+ \frac{\mu}{2} \log \left(\frac{\lvert K+K_{\tilde{\bv{Z}}_2} \rvert}{\lvert K+K_{\bv{Z}_2} \rvert} \right)\\
&= \frac{1}{2} \log \frac{\lvert K_{\bv{X}_1}+K_{\tilde{\bv{Z}}_1} \rvert}{ \lvert K_{\tilde{\bv{Z}}_1} \rvert }  - \frac{\mu}{2} \log \frac{\lvert K_{\bv{X}_1}+K_{\tilde{\bv{Z}}_2} \rvert}{\lvert K+K_{\tilde{\bv{Z}}_2} \rvert }+\frac{1}{2} \log \left( (2\pi e)^t |K_{\bv{Z}_1}| \right)- \frac{\mu}{2} \log \left((2\pi e )^t \lvert K+K_{\bv{Z}_2} \rvert \right).
\end{align*}
Now consider $ (K_{\bv{X}_1}+K_{\tilde{\bv{Z}}_1})^{-1} K_{\tilde{\bv{Z}}_1} $, we have
\begin{align*}
(K_{\bv{X}_1}+K_{\tilde{\bv{Z}}_1})^{-1} K_{\tilde{\bv{Z}}_1} 
&= (K_{\bv{X}_1}+K_{\tilde{\bv{Z}}_1})^{-1} (K_{\bv{X}_1}+K_{\tilde{\bv{Z}}_1}-K_{\bv{X}_1}) \\
&= I- (K_{\bv{X}_1}+K_{\tilde{\bv{Z}}_1})^{-1} K_{\bv{X}_1} \\
&\eqann[=]{a} I- \left((K_{\bv{X}_1}+K_{\bv{Z}_1})^{-1}+2M_1 \right) K_{\bv{X}_1} \\
&\eqann[=]{b} I- (K_{\bv{X}_1}+K_{\bv{Z}_1})^{-1} K_{\bv{X}_1} \\
&= I- (K_{\bv{X}_1}+K_{\bv{Z}_1})^{-1} (K_{\bv{X}_1}+K_{\bv{Z}_1}-K_{\bv{Z}_1}) \\
&= (K_{\bv{X}_1}+K_{\bv{Z}_1})^{-1} K_{\bv{Z}_1},
\end{align*}
where \eqannref{a} follows from \eqref{eq:enhancement_condition_Z1}, and \eqannref{b} follows from \eqref{eq:KKT2}.

Similarly, consider $ (K_{\bv{X}_1}+K_{\tilde{\bv{Z}}_2})^{-1} (K+K_{\tilde{\bv{Z}}_2}) $, we have
\begin{align*}
(K_{\bv{X}_1}+K_{\tilde{\bv{Z}}_2})^{-1} (K+K_{\tilde{\bv{Z}}_2})
&= (K_{\bv{X}_1}+K_{\tilde{\bv{Z}}_2})^{-1} (K_{\bv{X}_1}+K_{\tilde{\bv{Z}}_2}+K-K_{\bv{X}_1}) \\
&= I+ (K_{\bv{X}_1}+K_{\tilde{\bv{Z}}_2})^{-1} (K-K_{\bv{X}_1}) \\
&\eqann[=]{a} I+ \left((K_{\bv{X}_1}+K_{\bv{Z}_2})^{-1}+2\mu^{-1}M_2 \right) (K-K_{\bv{X}_1}) \\
&\eqann[=]{b} I+ (K_{\bv{X}_1}+K_{\bv{Z}_2})^{-1} (K-K_{\bv{X}_1}) \\
&= I+ (K_{\bv{X}_1}+K_{\bv{Z}_2})^{-1} (K+K_{\bv{Z}_2}-K_{\bv{X}_1}-K_{\bv{Z}_2}) \\
&= (K_{\bv{X}_1}+K_{\bv{Z}_2})^{-1} (K+K_{\bv{Z}_2}),
\end{align*}
where \eqannref{a} follows from \eqref{eq:enhancement_condition_Z2}, \eqannref{b} follows from \eqref{eq:KKT3}.

Thus
\begin{align}
(K_{\bv{X}_1}+K_{\tilde{\bv{Z}}_1})^{-1} K_{\tilde{\bv{Z}}_1} &= (K_{\bv{X}_1}+K_{\bv{Z}_1})^{-1} K_{\bv{Z}_1}, \\
(K_{\bv{X}_1}+K_{\tilde{\bv{Z}}_2})^{-1} (K+K_{\tilde{\bv{Z}}_2})
&= (K_{\bv{X}_1}+K_{\bv{Z}_2})^{-1} (K+K_{\bv{Z}_2}),
\end{align}
which implies that $ (P_G)=(\tilde{P}) $. This completes the proof of the conditional extremal inequality.

\bibliography{../bibliography}
\bibliographystyle{IEEEtran}

\end{document}